\documentclass[showpacs,twocolumn,superscriptaddress,aps]{revtex4}%
\usepackage{amsfonts}
\usepackage{amsmath}
\usepackage{amssymb}
\usepackage{epsfig}
\usepackage{graphicx}%
\ifx\pdfoutput\relax\let\pdfoutput=\undefined\fi
\newcount\msipdfoutput
\ifx\pdfoutput\undefined\else
\ifcase\pdfoutput\else
\ifx\paperwidth\undefined\else
\ifdim\paperheight=0pt\relax\else\pdfpageheight\paperheight\fi
\ifdim\paperwidth=0pt\relax\else\pdfpagewidth\paperwidth\fi
\fi\fi\fi
\begin{document}
\title{Bi-directional mapping between polarization and spatially encoded photonic qutrits}
\author{Qing Lin}
\email{qlin@mail.ustc.edu.cn}
\affiliation{College of Information Science and Engineering, Huaqiao University (Xiamen),
Xiamen 361021, China}
\author{Bing He}
\email{bhe98@earthlink.net}
\affiliation{Institute for Quantum Information Science, University of Calgary, Alberta T2N
1N4, Canada}

\pacs{03.67.Lx, 42.50.Ex}

\begin{abstract}
Qutrits, the triple level quantum systems in various forms, have been proposed
for quantum information processing recently. By the methods presented in this
paper a bi-photonic qutrit, which is encoded with the polarizations of two
photons in the same spatial-temporal mode, can be mapped to a single photon
qutrit in spatial modes. It will make arbitrary unitary operation on such
bi-photonic qutrit possible if we can also realize the inverse map to
polarization space. Among the two schemes proposed in this paper, the one
based only on linear optics realizes an arbitrary $U(3)$ operation with a very
small success probability. However, if added with weak nonlinearity, the
success probability can be greatly improved. These schemes are feasible with
the current experimental technology.

\end{abstract}
\maketitle

\section{Introduction}

Quantum communications and quantum computation apply quantum states to store
and transmit information. The capacity of a state for the purpose is dependent
on its dimension, so the higher dimension of a state means the higher capacity
to carry information. In addition, the use of higher dimensional quantum
states, e.g., qudits and entangled qudits, enjoys many advantages such as
enhanced security in quantum cryptography \cite{Langford}, more efficient
quantum logic gate \cite{Ralph} and others. Qudits and entangled qudits
therefore attract many researches recently. The proposals for generating
qudits and entangled qudits include orbital angular momentum entangled qutrits
\cite{Mair}, pixel entangled qudits \cite{Neves}, energy-time entangled and
time-bin entangled qudits \cite{Thew}, and bi-photonic qutrits encoded with
polarization degree of freedom \cite{Howell, Mikami, Lanyon}. In this paper we
will focus on bi-photonic qutrits which are represented with the polarizations
of two photons in the same spatial-temporal mode \cite{note}---$\left\vert
0\right\rangle _{3}\equiv\left\vert HH\right\rangle $, $\left\vert
1\right\rangle _{3}\equiv\left\vert HV\right\rangle $, and $\left\vert
2\right\rangle _{3}\equiv\left\vert VV\right\rangle $, where $H$ and $V$
denotes the horizontal and vertical polarization, respectively. The generation
of such qutrits including the entangled ones has been demonstrated
\cite{Howell, Lanyon}. In an recent work by Lanyon, \textit{et al.}
\cite{Lanyon}, with an ancilla qubit and a Fock state filter associated with
some wave plates, a bi-photonic state as the linear combination of
$\{\left\vert 0\right\rangle _{3},\left\vert 1\right\rangle _{3},\left\vert
2\right\rangle _{3}\}$ is generated from the logic state $\left\vert
0\right\rangle _{3}$. To manipulate a bi-photonic qutrit in this form, one
should know how to implement a unitary operation on such qutrits. However, due
to the indistinguishableness of two photons in the same spatial-temporal mode,
it is very difficult to realize a simple unitary operation on such bi-photonic
qutrit \cite{Bogdanov}. Here we present two schemes realizing the
transformation from a bi-photonic qutrit to any other bi-photonic qutrit, i.
e., arbitrary unitary operations $U(3)$ on bi-photonic qutrits. The schemes
work with transforming the input bi-photonic qutrits to the corresponding
single photon qutrits in spatial modes, and then mapping the single photon
qutrits back to the original polarization modes of two photons.

The rest of the paper is organized as follows. In Sec. II, we present a purely
linear optical scheme of the transformation and inverse transformation from a
bi-photonic qutrit in the same spatial-temporal mode to the corresponding
single photon qutrit. In Sec. III, we improve on the linear optical scheme
with weak cross-Kerr nonlinearity, making the realization of bi-directional
mapping much more efficient. Sec. IV concludes the work with a brief discussion.

\section{\bigskip Bi-directional mapping with linear optical elements}

Any unitary operation on a single photon qudits in spatial modes can be
performed by a linear optical multi-port interferometer (LOMI) \cite{Reck}. It
is therefore possible to manipulate bi-photonic qutrits following such
strategy---first transforms a bi-photonic qutrit to a single-photon qutrit,
and then performs the desired operations on this single photon qutrit, and
finally transforms the single photon qutrit back to a bi-photonic qutrit. In
what follows, we present the details of the procedure, which is realized only
with linear optical elements.

\subsection{\bigskip Transforming bi-photon qutrit to single-photon qutrit}

\begin{figure}[tbh]
\begin{center}
\epsfig{file=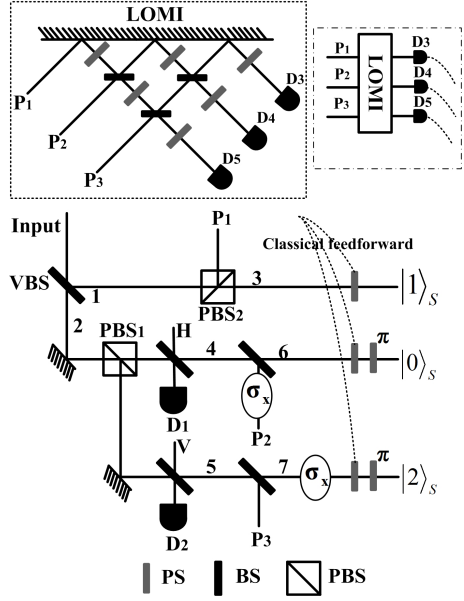,width=6cm}
\end{center}
\caption{Schematic setup for the transformation from a bi-photonic qutrit to
the corresponding single-photon qutrit. At first, the input qutrit is
transmitted through a VBS, and then the two output modes are transmitted
through a PBS, respectively. Two single photons are used as the ancillas,
which will interfere with the output modes of PBS$_{1}$. The part in dashed
line is used to erase the path information of the modes $P_{1},P_{2},P_{3}$,
which are all in the state $\left\vert V\right\rangle $. The detection results
are used as control signals of the conditional phase shift summarized in Tab.
I through the classical feed-forward. By the proper post-selection, the
bi-photonic qutrit can be transformed to the corresponding single-photon
qutrit as in Eq. (\ref{2}). For details, see the text.}%
\end{figure}

Suppose a bi-photonic qutrit is initially prepared as
\begin{align}
\left\vert \psi\right\rangle _{in}=\alpha\left\vert 0\right\rangle _{3}%
+\beta\left\vert 1\right\rangle _{3}+\gamma\left\vert 2\right\rangle _{3},
\label{1}%
\end{align}
where $\left\vert \alpha\right\vert ^{2}+\left\vert \beta\right\vert
^{2}+\left\vert \gamma\right\vert ^{2}=1$. The operations shown in Fig. 1
implements the map
\begin{align}
\left\vert \psi\right\rangle _{in}\rightarrow\alpha\left\vert 0\right\rangle
_{S}+\beta\left\vert 1\right\rangle _{S}+\gamma\left\vert 2\right\rangle
_{S}=\left\vert \psi\right\rangle _{S}, \label{2}%
\end{align}
where $\left\vert \psi\right\rangle _{S}$ is a single-photon qutrit encoded
with the spatial modes $|i\rangle_{S}$ ($i=0,1,2$) of the single photon. Here
we first apply a variable beam splitter (VBS) to the input bi-photonic qutrit,
realizing the following transformation%

\begin{align}
&  \left(  \frac{\alpha}{\sqrt{2}}a_{H}^{\dagger2}+\beta a_{H}^{\dagger}%
a_{V}^{\dagger}+\frac{\gamma}{\sqrt{2}}a_{V}^{\dagger2}\right)  \left\vert
vac\right\rangle \nonumber\\
&  \rightarrow\left[  \frac{\alpha}{\sqrt{2}}\left(  ra_{H1}^{\dagger}%
+ta_{H2}^{\dagger}\right)  ^{2}+\beta\left(  ra_{H1}^{\dagger}+ta_{H2}%
^{\dagger}\right)  \right.  \nonumber\\
&  \times\left.  \left(  ra_{V1}^{\dagger}+ta_{V2}^{\dagger}\right)
+\frac{\gamma}{\sqrt{2}}\left(  ra_{V1}^{\dagger}+ta_{V2}^{\dagger}\right)
^{2}\right]  \left\vert vac\right\rangle ,
\end{align}
where $1$, $2$ denote the different paths, and $t$ $(r)$ is the transmissivity
(reflectivity) of the VBS.

Next, in order to project out the proper components, we introduce two single
photons $\left\vert H\right\rangle $ and $\left\vert V\right\rangle $ as the
ancilla, and make them interfere with the output modes of the polarizing beam
splitters (PBS$_{1}$) on two 50:50 beam splitters (BS), respectively. Due to
the Hong-Ou-Mandal interference effect \cite{Hong}, two indistinguishable
photon will be bunching to the same output mode of BS, and then we can use the
proper post-selection to get the desired components. To see the details, we
show the evolution of each input mode of two photons as follows:%

\begin{align}
a_{H1}^{\dagger} &  \rightarrow a_{H3}^{\dagger},a_{H2}^{\dagger}%
\rightarrow\frac{1}{\sqrt{2}}\left(  a_{H4}^{\dagger}+a_{HD_{1}}^{\dagger
}\right)  ,\nonumber\\
a_{V1}^{\dagger} &  \rightarrow a_{VP_{1}}^{\dagger},a_{V2}^{\dagger
}\rightarrow\frac{1}{\sqrt{2}}\left(  a_{V5}^{\dagger}+a_{VD_{2}}^{\dagger
}\right)  ,
\end{align}
where the subscripts $D_{1},D_{2}$ denote the modes going to photon number
non-resolving detectors. Meanwhile, for the ancilla photons, the evolutions
are
\begin{align}
a_{H}^{\dagger} &  \rightarrow\frac{1}{\sqrt{2}}\left(  a_{H4}^{\dagger
}-a_{HD_{1}}^{\dagger}\right)  ,\ \nonumber\\
a_{V}^{\dagger} &  \rightarrow\frac{1}{\sqrt{2}}\left(  a_{V5}^{\dagger
}-a_{VD_{2}}^{\dagger}\right)  .
\end{align}
The 50:50 BS placed on path 4 (5) is to split the mode 4 (5) into two output
modes 6, P$_{2}$ (7, P$_{3}$), making the transformations,
\begin{align}
a_{H4}^{\dagger} &  \rightarrow\frac{1}{\sqrt{2}}\left(  a_{H6}^{\dagger
}+a_{HP_{2}}^{\dagger}\right)  \rightarrow\frac{1}{\sqrt{2}}\left(
a_{H6}^{\dagger}+a_{VP_{2}}^{\dagger}\right)  ,\nonumber\\
a_{V5}^{\dagger} &  \rightarrow\frac{1}{\sqrt{2}}\left(  a_{V7}^{\dagger
}+a_{VP_{3}}^{\dagger}\right)  \rightarrow\frac{1}{\sqrt{2}}\left(
a_{H7}^{\dagger}+a_{VP_{3}}^{\dagger}\right)  .
\end{align}
After that, one obtains the following state:%

\begin{align}
&  \left(  -\frac{\alpha}{4\sqrt{2}}t^{2}a_{H6}^{\dagger}a_{VP_{2}}^{\dagger
}+\frac{\beta}{2}r^{2}a_{H3}^{\dagger}a_{VP_{1}}^{\dagger}-\frac{\gamma
}{4\sqrt{2}}t^{2}a_{V7}^{\dagger}a_{VP_{3}}^{\dagger}\right) \nonumber\\
&  \times a_{HD_{1}}^{\dagger}a_{VD_{2}}^{\dagger}\left\vert vac\right\rangle
+rest. \label{sq}%
\end{align}
where $rest.$ denotes the components of two photons appearing in the same
spatial mode. If we discard the modes $P_{1},P_{2},P_{3}$ without changing
anything else, i.e., erase the path information of $P_{1},P_{2},P_{3}$, the
first three terms in Eq. (\ref{sq}) will be just the desired single photon
qutrit, which carries the same coefficients of the input bi-photon qutrit.
Since there is only one photon in the modes $P_{1},P_{2},P_{3}$, we will use a
quantum Fourier transform (QFT) ($j,k^{\prime}$ denote the spatial modes)
\cite{Nielsen},
\begin{equation}
a_{Vj}^{\dagger}\left\vert vac\right\rangle =\frac{1}{\sqrt{3}}\overset
{2}{\underset{k^{\prime}=0}{{\sum}}}e^{2\pi ijk/3}a_{Vk^{\prime}}^{\dagger
}\left\vert vac\right\rangle , \label{qft}%
\end{equation}
to do it. The QFT is an unitary operation for a single photon in three spatial
modes, so we can use an LOMI shown in the dashed line of Fig. 1 to implement
it. Just like the setups in the dash-dotted line, three photon number
non-resolving detectors are used and the detection results are to control the
conditional phase shift (PS) through classical feedforward. The relations
between the detection results and the corresponding PS operations are
summarized in Tab. \ref{tb1}. \begin{table}[ptb]
$%
\begin{array}
[c]{|c|c|c|c|}\hline
& D_{3} & D_{4} & D_{5}\\\hline
3 & 0 & 0 & 0\\\hline
6 & 0 & \frac{2\pi}{3} & \frac{4\pi}{3}\\\hline
7 & 0 & \frac{4\pi}{3} & \frac{8\pi}{3}\\\hline
\end{array}
$\caption{The relations between the detections and the corresponding phase
shifters on path 3, 6, 7.}%
\label{tb1}%
\end{table}After that, with the coincident measurements of the detectors
$D_{1}$, $D_{2}$, and one of the detectors $D_{3},D_{4},D_{5}$, the state%

\begin{equation}
\left(  \frac{\alpha}{4\sqrt{2}}t^{2}a_{H6}^{\dagger}+\frac{\beta}{2}%
r^{2}a_{H3}^{\dagger}+\frac{\gamma}{4\sqrt{2}}t^{2}a_{V7}^{\dagger}\right)
\left\vert vac\right\rangle ,
\end{equation}
will be projected out by the post-selection. We can rewrite it as
\begin{equation}
\frac{\alpha}{4\sqrt{2}}t^{2}\left\vert H\right\rangle _{6}+\frac{\beta}%
{2}r^{2}\left\vert H\right\rangle _{3}+\frac{\gamma}{4\sqrt{2}}t^{2}\left\vert
V\right\rangle _{7}.
\end{equation}
It is straightforward that the state will be $\left\vert \psi\right\rangle
_{S}$, given that $t^{2}=2\sqrt{2}r^{2}$ or $t^{2}=\frac{2\sqrt{2}}%
{1+2\sqrt{2}}$. The corresponding success probability of the process is
$\left(  \frac{t^{2}}{4\sqrt{2}}\right)  ^{2}=1.71\times10^{-2}.$

\subsection{Transformation back to bi-photonic qutrit}

\begin{figure}[tbh]
\begin{center}
\epsfig{file=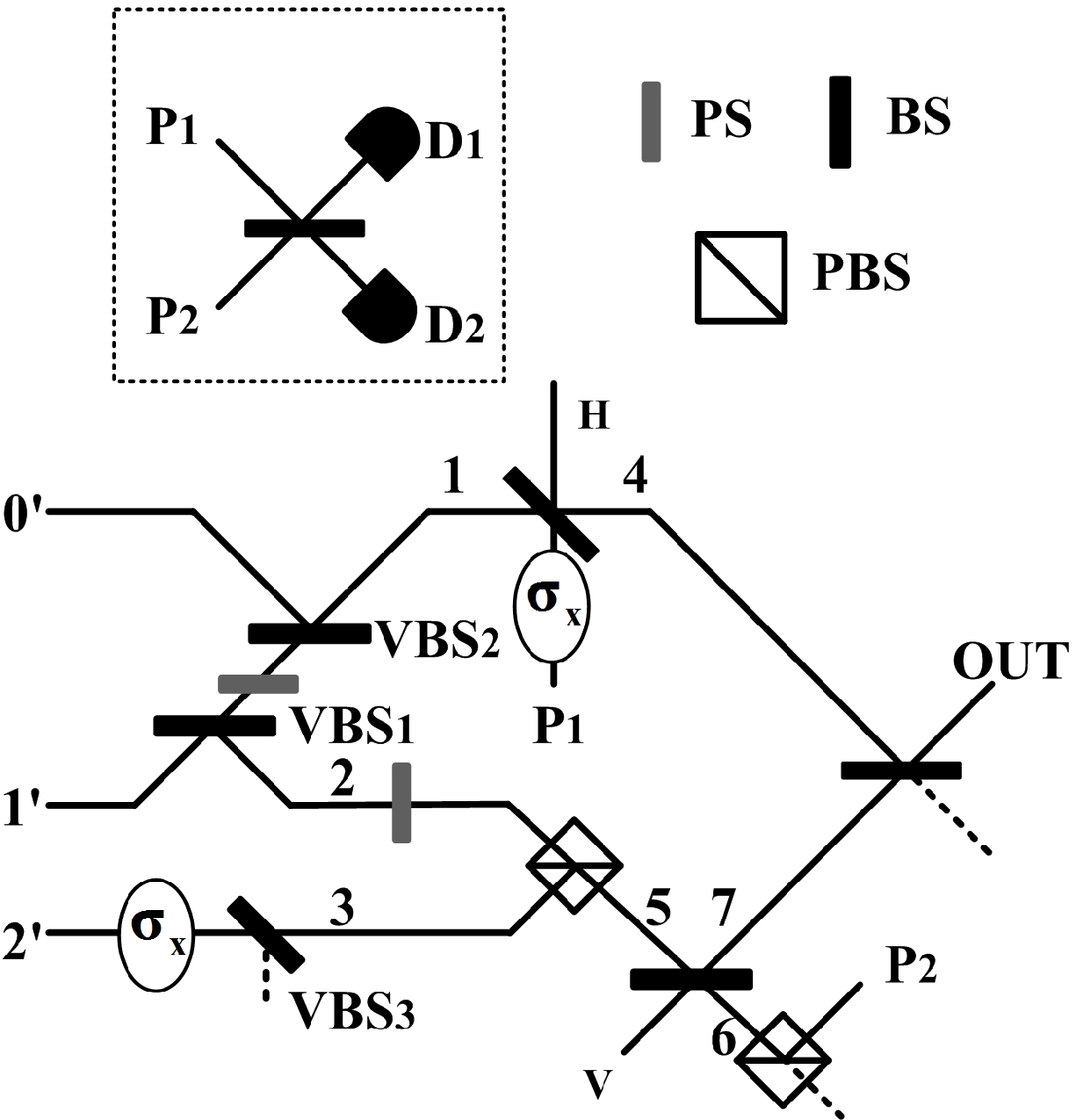,width=6cm}
\end{center}
\caption{Schematic setup for the inverse transformation from a single-photon
qutrit back to a bi-photon qutrit. Two variable beam splitters (VBS) are
applied, and two extra single photon work as the ancilla. The single-photon
qutrit can be transformed back to a bi-photonic qutrit by post-selection. For
details, see the text.}%
\end{figure}

After the desired operations performed on the single photon qutirt, we should
transform the single-photon qutrit $\left\vert \psi^{^{\prime}}\right\rangle
_{S}=\alpha^{^{\prime}}\left\vert 0^{^{\prime}}\right\rangle _{S}%
+\beta^{^{\prime}}\left\vert 1^{^{\prime}}\right\rangle _{S}+\gamma^{^{\prime
}}\left\vert 2^{\prime}\right\rangle _{S}$ back to a bi-photonic qutrit. The
inverse transformation is shown in Fig. 2. Three VBSs (VBS$_{1}$, VBS$_{2}$,
VBS$_{3}$) with the transmissivities (reflectivities) $t_{1}$, $t_{2}$,
$t_{3}$ ($r_{1}$, $r_{2}$, $r_{3}$), two PSs and $\sigma_{x}$ operation are
applied to perform the transformation%

\begin{align}
&  \left(  \alpha^{^{\prime}}a_{H0^{^{\prime}}}^{\dagger}+\beta^{^{\prime}%
}a_{H1^{^{\prime}}}^{\dagger}+\gamma^{^{\prime}}a_{H2^{^{\prime}}}^{\dagger
}\right)  \left\vert vac\right\rangle \nonumber\\
&  \rightarrow\left[  \left(  \alpha^{^{\prime}}r_{2}-\beta^{^{\prime}}%
t_{1}t_{2}\right)  a_{H1}^{\dagger}+\beta^{^{\prime}}r_{1}a_{H2}^{\dagger
}+\gamma^{^{\prime}}t_{3}a_{V3}^{\dagger}\right]  \left\vert vac\right\rangle
.
\end{align}
In order to select out the desired components, we introduce a single photon
$\left\vert H\right\rangle $\ as an ancilla, which will interfere with the
mode 1 through a 50:50 BS, and meanwhile combine the modes 2 and 3 by a PBS
into the mode 5, which will then interfere with another ancilla single photon
$\left\vert V\right\rangle $ through a 50:50 BS. The total state will be then
transformed to%
\begin{align}
&  \left\{  \frac{1}{2\sqrt{2}}\left(  \alpha^{^{\prime}}r_{2}-\beta
^{^{\prime}}t_{1}t_{2}\right)  \left(  a_{H4}^{\dagger2}-a_{HP_{1}}^{\dagger
2}\right)  \left(  a_{V7}^{\dagger}-a_{V6}^{\dagger}\right)  \right.
\nonumber\\
&  +\frac{1}{2\sqrt{2}}\beta^{^{\prime}}r_{1}\left(  a_{H4}^{\dagger
}-a_{HP_{1}}^{\dagger}\right)  \left(  a_{H7}^{\dagger}+a_{H6}^{\dagger
}\right)  \left(  a_{V7}^{\dagger}-a_{V6}^{\dagger}\right) \nonumber\\
&  \left.  +\frac{1}{2\sqrt{2}}\gamma^{^{\prime}}t_{3}\left(  a_{H4}^{\dagger
}-a_{HP_{1}}^{\dagger}\right)  \left(  a_{V7}^{\dagger2}-a_{V6}^{\dagger
2}\right)  \right\}  \left\vert vac\right\rangle .
\end{align}
The $\left\vert V\right\rangle $ mode on path 6 will be reflected to mode
$P_{2}$. Now, the following state can be achieved:
\begin{align}
&  -\frac{1}{2\sqrt{2}}\left[  \left(  \alpha^{^{\prime}}r_{2}-\beta
^{^{\prime}}t_{1}t_{2}\right)  a_{H4}^{\dagger2}a_{VP_{2}}^{\dagger}\right.
\nonumber\\
&  +\beta^{^{\prime}}r_{1}\left(  a_{H7}^{\dagger}a_{V7}^{\dagger}a_{HP_{1}%
}^{\dagger}+a_{H4}^{\dagger}a_{H7}^{\dagger}a_{VP_{2}}^{\dagger}\right)
\nonumber\\
&  \left.  +\gamma^{^{\prime}}t_{3}a_{V7}^{\dagger2}a_{HP_{1}}^{\dagger
}\right]  \left\vert vac\right\rangle +rest.
\end{align}
The left work will be the erasure of the path information of the modes $P_{1},
P_{2}$ by the detection similar to that in II. A. Because there are only two
spatial modes, the realization of the QFT will be simplified with just one
50:50 BS as shown in dashed line. Now, the state%
\begin{align}
&  -\frac{1}{4}\left[  \left(  \alpha^{^{\prime}}r_{2}-\beta^{^{\prime}}%
t_{1}t_{2}\right)  a_{H4}^{\dagger2}\right. \nonumber\\
&  +\beta^{^{\prime}}r_{1}\left(  a_{H7}^{\dagger}a_{V7}^{\dagger}%
+a_{H4}^{\dagger}a_{H7}^{\dagger}\right) \nonumber\\
&  \left.  +\gamma^{^{\prime}}t_{3}a_{V7}^{\dagger2}\right]  a_{VD_{1}%
}^{\dagger}\left\vert vac\right\rangle +rest.,
\end{align}
can be achieved, where we only keep the terms with the photonic modes on
$P_{1},P_{2}$ being detected by the detector $D_{1}$. In the other case when
the photon is detected by the detector $D_{2}$, there will be an additional
phase shift $\pi$ to the components including $a_{VP_{2}}^{\dagger}$, and it
seems difficult to remove it by a simple operation.

After the erasure of $P_{1}$ and $P_{2}$ modes, the modes 4 and 7 will
interfere with each other through a 50:50 BS. If there are two photons in the
final output (which can be realized by common bi-photonic qutrit tomograph
\cite{Lanyon}) and a click on one of the two detectors $D_{1}$, $D_{2}$, we
will project out the state
\begin{align}
&  -\frac{1}{8}\left[  \left(  \alpha^{^{\prime}}r_{2}-\beta^{^{\prime}}%
t_{1}t_{2}\right)  a_{Hout}^{\dagger2}\right. \nonumber\\
&  \left.  +\beta^{^{\prime}}r_{1}\left(  a_{Hout}^{\dagger}a_{Vout}^{\dagger
}+a_{Hout}^{\dagger2}\right)  +\gamma^{^{\prime}}t_{3}a_{Vout}^{\dagger
2}\right]  \left\vert vac\right\rangle \nonumber\\
&  =-\frac{1}{8}\left[  \left(  \alpha^{^{\prime}}r_{2}-\beta^{^{\prime}}%
t_{1}t_{2}+\beta^{^{\prime}}r_{1}\right)  a_{Hout}^{\dagger2}\right.
\nonumber\\
&  \left.  +\beta^{^{\prime}}r_{1}a_{Hout}^{\dagger}a_{Vout}^{\dagger}%
+\gamma^{^{\prime}}t_{3}a_{Vout}^{\dagger2}\right]  \left\vert
vac\right\rangle
\end{align}
by post-selection. Choosing $t_{1}t_{2}=r_{1}$ and $\sqrt{2}r_{2}=$\ $r_{1}$,
i.e., $t_{1}^{2}=\frac{\sqrt{17}-3}{2}$ ($r_{1}^{2}=\frac{5-\sqrt{17}}{2}$),
associated with $t_{3}^{2}=\frac{5-\sqrt{17}}{4},$ we can achieve the final
state
\begin{equation}
\frac{r_{1}}{8}\left(  \frac{\alpha^{^{\prime}}}{\sqrt{2}}a_{Hout}^{\dagger
2}+\beta^{^{\prime}}a_{Hout}^{\dagger}a_{Vout}^{\dagger}+\frac{\gamma
^{^{\prime}}}{\sqrt{2}}a_{Vout}^{\dagger2}\right)  \left\vert vac\right\rangle
,
\end{equation}
which is the target bi-photonic qutrit $\alpha^{^{\prime}}\left\vert
0\right\rangle _{3}+\beta^{^{\prime}}\left\vert 1\right\rangle _{3}%
+\gamma^{^{\prime}}\left\vert 2\right\rangle _{3}$. The corresponding success
probability is $\left(  \frac{r_{1}}{8}\right)  ^{2}=6.85\times10^{-3}$.
Assocaited with the above transformation, we could manipulate the bi-photonic
qutrits, such as perform an arbitrary unitary operation $U(3)$ on them, with a
success probability $1.71\times6.85\times10^{-5}=1.17\times10^{-4}$. The
scheme succeeds with a very small probability, but in principle it can realize
any unitary operation on a bi-photon qutrit.

In summary, with four ancilla single photons, we could realize arbitrary
manipulation with linear optical elements and coincidence measurements. Since
only two cases--no photon or any number of photons---should be discriminated,
the common photon number non-resolving detector, e.g., silicon avalanche
photodiodes (APDs) will be necessary for the scheme.

\section{Bi-directional mapping with weak cross-Kerr nonlinearity}

\label{sec3}

The success probability of the above scheme with only linear optical elements
could be too small for practical application. This success probability,
however, can be greatly increased if we apply some weak nonlinearity in the
circuit. The application of weak cross-Kerr nonlinearity has been proposed in
various fields of quantum information science. It was firstly applied to
realize parity projector \cite{Barrett} and deterministic CNOT gate
\cite{Nemoto}, and then in some quantum computation and communication schemes
(see, e.g., \cite{Spiller, Lin, He}). The effective Hamiltonian for cross-Kerr
nonlinearity is $\mathcal{H}=-\hbar\chi\hat{n}_{i}\hat{n}_{j}$ ($\chi$ is the
nonlinear intensity and $\hat{n}_{i/j}$ the number operator of the interacting
modes). The cross phase modulation (XPM) process caused by such interaction
between a Fock state $|n\rangle$ and a coherent state $|\alpha\rangle$ gives
rise to the transformation, $|n\rangle|\alpha\rangle\rightarrow|n\rangle
|\alpha e^{i\theta}\rangle$, where $\theta=\chi t$ induced during the
interaction time $t$ could be small with weak nonlinearity. Another useful
technique to our schem is homodyne-heterodyne measurement for the quadratures
of coherent state. A state like $\sum_{k}|k\rangle|\alpha e^{ik\theta}\rangle$
can be projected to a definite Fock state or a superposition of some Fock
states by such measurement, which can be performed with high fidelity.

\subsection{\bigskip Transformation with XPM process}

\begin{figure}[tbh]
\begin{center}
\epsfig{file=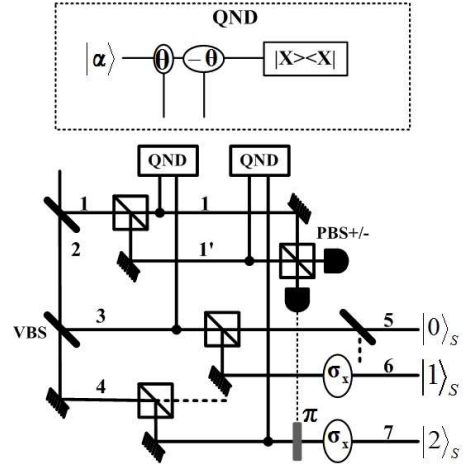,width=6cm}
\end{center}
\caption{Schematic setup for the transformation with XPM process. Two QND
modules working with XPM process are used here. The transformation is realized
under the condition that the two qubus beams pick no phase shift, with the
corresponding success probability 1/6. For the details, see text.}%
\end{figure}

With weak cross-Kerr nonlinearity, we implement the transformation from a
bi-photonic qutrit to a single photon qutrit as shown in Fig. 3. An initial
bi-photonic qutrit in the state $\left\vert \psi\right\rangle _{in}$ of Eq.
(\ref{1}) is first sent to a 50:50 BS, making the following transformation
\begin{align}
&  \left(  \frac{\alpha}{\sqrt{2}}a_{H}^{\dagger2}+\beta a_{H}^{\dagger}%
a_{V}^{\dagger}+\frac{\gamma}{\sqrt{2}}a_{V}^{\dagger2}\right)  \left\vert
vac\right\rangle \nonumber\\
&  \rightarrow\left[  \frac{\alpha}{2\sqrt{2}}\left(  a_{H1}^{\dagger}%
+a_{H2}^{\dagger}\right)  ^{2}+\frac{\beta}{2}\left(  a_{H1}^{\dagger}%
+a_{H2}^{\dagger}\right)  \right. \nonumber\\
&  \left.  \times\left(  a_{V1}^{\dagger}+a_{V2}^{\dagger}\right)
+\frac{\gamma}{2\sqrt{2}}\left(  a_{V1}^{\dagger}+a_{V2}^{\dagger}\right)
^{2}\right]  \left\vert vac\right\rangle . \label{die}%
\end{align}
Next a VBS is placed on path 2 such that
\begin{align}
a_{H2}^{\dagger}  &  \rightarrow ra_{H3}^{\dagger}+ta_{H4}^{\dagger
},\nonumber\\
a_{V2}^{\dagger}  &  \rightarrow ra_{V3}^{\dagger}+ta_{V4}^{\dagger}.
\end{align}
Then, after three PBSs change the spatial modes as $1\rightarrow1,1^{\prime}$,
$3\rightarrow5,6$, $4\rightarrow7$, two qubus beams (i.e. coherent states)
$\left\vert \alpha_{1}\right\rangle \left\vert \alpha_{2}\right\rangle $ will
be coupled to the corresponding photonic modes through the XPM processes in
two quantum nondemolition detection (QND) modules, which are shown in dashed
line of Fig. 3. The result will be the following transformation of the total
system:
\begin{align}
&  \left(  \frac{\alpha}{\sqrt{2}}ra_{H1}^{\dagger}a_{H5}^{\dagger}%
+\frac{\beta}{2}ra_{H1}^{\dagger}a_{V6}^{\dagger}+\frac{\gamma}{\sqrt{2}%
}ta_{V1^{\prime}}^{\dagger}a_{V7}^{\dagger}\right)  \left\vert
vac\right\rangle \nonumber\\
&  \times\left\vert \alpha_{1}\right\rangle \left\vert \alpha_{2}\right\rangle
+rest.,
\end{align}
where we only give the terms that two qubus beams pick no phase shift. These
terms can be separated from the others by the quadrature measurement
$\left\vert X\right\rangle \left\langle X\right\vert $, which is implementable
with homodyne-heterodyne measurement \cite{Nemoto,Spiller}, to obtain the
following state
\begin{align}
&  \left(  \frac{\alpha}{\sqrt{2}}ra_{H1}^{\dagger}a_{H5}^{\dagger}%
+\frac{\beta}{2}ra_{H1}^{\dagger}a_{V6}^{\dagger}+\frac{\gamma}{\sqrt{2}%
}ta_{V1^{\prime}}^{\dagger}a_{V7}^{\dagger}\right)  \left\vert
vac\right\rangle \nonumber\\
&  =\frac{\alpha}{\sqrt{2}}r\left\vert H\right\rangle _{1}\left\vert
H\right\rangle _{5}+\frac{\beta}{2}r\left\vert H\right\rangle _{1}\left\vert
V\right\rangle _{6}+\frac{\gamma}{\sqrt{2}}t\left\vert V\right\rangle
_{1}\left\vert V\right\rangle _{7}. \label{kuie}%
\end{align}
This state can be expressed as
\begin{align}
&  \frac{\alpha}{2}r\left(  \left\vert +\right\rangle _{1}+\left\vert
-\right\rangle _{1}\right)  \left\vert H\right\rangle _{5}+\frac{\beta}%
{2\sqrt{2}}r\left(  \left\vert +\right\rangle _{1}+\left\vert -\right\rangle
_{1}\right)  \left\vert V\right\rangle _{6}\nonumber\\
&  +\frac{\gamma}{2}t\left(  \left\vert +\right\rangle _{1}-\left\vert
-\right\rangle _{1}\right)  \left\vert V\right\rangle _{7},
\end{align}
where $\left\vert \pm\right\rangle =\frac{1}{\sqrt{2}}\left(  \left\vert
H\right\rangle +\left\vert V\right\rangle \right)  $. Now, we use a PBS$_{\pm
}$ which transmits $\left\vert +\right\rangle $ and reflects $\left\vert
-\right\rangle $, and the following two photon number non-resolving detectors.
If the detection is $\left\vert +\right\rangle $, the state%
\begin{equation}
\frac{\alpha}{\sqrt{2}}r\left\vert H\right\rangle _{5}+\frac{\beta}%
{2}r\left\vert V\right\rangle _{6}+\frac{\gamma}{\sqrt{2}}t\left\vert
V\right\rangle _{7} \label{22}%
\end{equation}
will be projected out; on the other hand, if the detection is $\left\vert
-\right\rangle $, what is realized is
\begin{equation}
\frac{\alpha}{\sqrt{2}}r\left\vert H\right\rangle _{5}+\frac{\beta}%
{2}r\left\vert V\right\rangle _{6}-\frac{\gamma}{\sqrt{2}}t\left\vert
V\right\rangle _{7},
\end{equation}
which can be transform to the state in Eq. (\ref{22}) by the conditional phase
shifter $\pi$ on path 7. By selecting $\frac{r}{2}=\frac{t}{\sqrt{2}}$, i.e.,
$t=\frac{1}{\sqrt{3}}$, and using a 50:50 BS for the mode 5 and two
$\sigma_{x}$ operations for modes 6, 7, we can achieve the following state,
\begin{equation}
\frac{1}{\sqrt{6}}\left(  \alpha\left\vert H\right\rangle _{5}+\beta\left\vert
H\right\rangle _{6}+\gamma\left\vert H\right\rangle _{7}\right)  ,
\end{equation}
which is the single photon qutrit $\left\vert \psi\right\rangle _{S}$ in Eq.
(\ref{2}). The success probability is $\frac{1}{6}$, which is much higher than
that of the linear optical scheme. Moreover, no ancilla single photon is
necessary here.

The scheme is based on quadrature projection after XPM process, so it does not
require any post-selection by coincidence measurement. But it needs an XPM
phase shift of $-\theta$, which is only possible with the equivalent phase
shift $2\pi-\theta$ and could be impractical \cite{Kok}. The XPM phase shift
$\theta\sim10^{-2}$ is possible with, for example, electromagnetically induced
transparencies (EIT) \cite{EIT}, whispering-gallery microresonators
\cite{WGM}, optical fibers \cite{OF}, or cavity QED systems \cite{QED}, but
the corresponding $2\pi-\theta$ will be too large to realize by the available
techniques. To avoid the XPM phase shift of $-\theta$, we propose a different
design of the transformation shown in Fig. 4. Here we use the double XPM
method in \cite{He} to replace the two XPM processes without changing anything else.

\begin{figure}[tbh]
\begin{center}
\epsfig{file=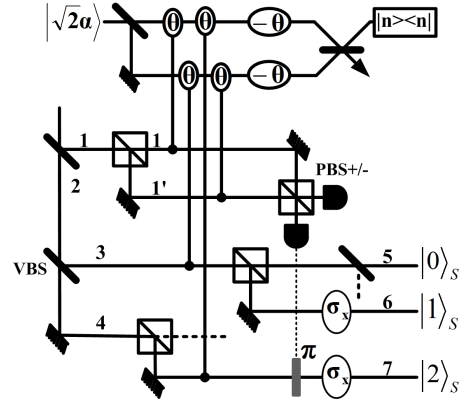,width=6cm}
\end{center}
\caption{Schematic setup for the transformation from bi-photon qutrits to the
corresponding single photon qutrits with double XPM method. The only
difference from Fig. 3 is that the two separate XPM processes are replaced by
a double XPM process of two identical qubus beams. In the design, no XPM phase
shift of $-\theta$ will be necessary, and it makes the scheme more feasible.}%
\end{figure}

We describe it briefly as the process is similar. In the double XPM process
two qubus beams $\left\vert \alpha\right\rangle \left\vert \alpha\right\rangle
$ will be coupled to the corresponding photonic modes as shown in Fig. 4. The
XPM pattern in Fig. 4 is that the first beam being coupled to the $\left\vert
H\right\rangle $ mode on path 1 and the $\left\vert V\right\rangle $ mode on
path 4, while the second beam to $\left\vert V\right\rangle $ mode on path
$1^{\prime}$ and the modes on path 3. Suppose the XPM phase shifts induced by
the couplings are all $\theta$. After that, the total system will be
transformed to
\begin{align}
&  \left(  \frac{\alpha}{\sqrt{2}}ra_{H1}^{\dagger}a_{H3}^{\dagger}%
+\frac{\beta}{2}a_{H1}^{\dagger}a_{V1^{\prime}}^{\dagger}+\frac{\beta}%
{2}ra_{H1}^{\dagger}a_{V3}^{\dagger}\right. \nonumber\\
&  \left.  +\frac{\beta}{2}rta_{H3}^{\dagger}a_{V4}^{\dagger}+\frac{\gamma
}{2\sqrt{2}}ta_{V1}^{\dagger}a_{V4}^{\dagger}+\frac{\gamma}{2\sqrt{2}}%
rta_{V3}^{\dagger}a_{V4}^{\dagger}\right) \nonumber\\
&  \times\left\vert vac\right\rangle \left\vert \alpha e^{i\theta
}\right\rangle \left\vert \alpha e^{i\theta}\right\rangle +\frac{\alpha
}{2\sqrt{2}}t^{2}a_{H4}^{\dagger2}\left\vert vac\right\rangle \left\vert
\alpha\right\rangle \left\vert \alpha\right\rangle +rest.,
\end{align}
where $rest.$ denotes the terms that the two qubus beams pick up the different
phase shifts. A phase shifter of $-\theta$ is respectively applied to two
qubus beams, and then one more 50:50 BS implements the transformation
$\left\vert \alpha_{1}\right\rangle \left\vert \alpha_{2}\right\rangle
\rightarrow\left\vert \frac{\alpha_{1}-\alpha_{2}}{\sqrt{2}}\right\rangle
\left\vert \frac{\alpha_{1}+\alpha_{2}}{\sqrt{2}}\right\rangle $ of the
coherent-state components. The above state will be therefore transformed to
\begin{align}
&  \left(  \frac{\alpha}{\sqrt{2}}ra_{H1}^{\dagger}a_{H3}^{\dagger}%
+\frac{\beta}{2}a_{H1}^{\dagger}a_{V1^{\prime}}^{\dagger}+\frac{\beta}%
{2}ra_{H1}^{\dagger}a_{V3}^{\dagger}\right. \nonumber\\
&  \left.  +\frac{\beta}{2}rta_{H3}^{\dagger}a_{V4}^{\dagger}+\frac{\gamma
}{2\sqrt{2}}ta_{V1}^{\dagger}a_{V4}^{\dagger}+\frac{\gamma}{2\sqrt{2}}%
rta_{V3}^{\dagger}a_{V4}^{\dagger}\right) \nonumber\\
&  \times\left\vert vac\right\rangle \left\vert 0\right\rangle \left\vert
\sqrt{2}\alpha\right\rangle +\frac{\alpha}{2\sqrt{2}}t^{2}a_{H4}^{\dagger
2}\left\vert vac\right\rangle \left\vert 0\right\rangle \left\vert \sqrt
{2}\alpha\right\rangle +rest.
\end{align}
Then, we could use the projections $\left\vert n\right\rangle \left\langle
n\right\vert $ on the first qubus beam to get the proper output. If $n=0$, and
by the post-selection that one photon will appear on the output (5, 6, 7)
while a click on one of the two detectors after the PBS$_{\pm}$, the state in
Eq. (\ref{kuie}) can be therefore projected out. Similar to the process in
Fig. 3, we can achieve the final single photon qutrit $\left\vert
\psi\right\rangle _{S}$ with the success probability $\frac{1}{6}.$ Though
this design requires the post-selection, it dispenses with the XPM phase shift
of $-\theta$, so it could be more experimentally feasible.

\subsection{Inverse transformation with XPM process}

\begin{figure}[tbh]
\begin{center}
\epsfig{file=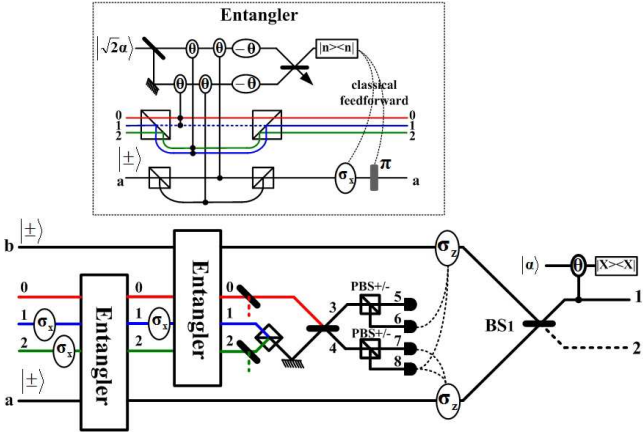,width=8.5cm}
\end{center}
\caption{Schematic setup for the inverse transformation with XPM process. The
part called Entangler shown in dashed line entangles the single photon qutrit
and the ancilla single photon. Out of the Entangler, the polarization of the
ancilla photon will be the same as those of the single photon qutrit. This
inverse transformation can be implemented with a success probability 1/2.}%
\end{figure}

Now we should transform the output single photon qutrit back to a bi-photonic
qutrit. We apply the inverse transformation procedure shown in Fig. 5, and
will show that it can be realized with a success probability as high as
$\frac{1}{2}$.

At first, we apply a setup called Entangler shown in dashed line to the
transformed single photon qutrit $\left\vert \psi^{^{\prime}}\right\rangle
_{S}=\alpha^{^{\prime}}\left\vert H\right\rangle _{0}+\beta^{^{\prime}%
}\left\vert H\right\rangle _{1}+\gamma^{^{\prime}}\left\vert H\right\rangle
_{2}$ with an ancilla single photon in the state $\left\vert \pm\right\rangle
_{a}$, after a $\sigma_{x}$ operation performed on the spatial modes $1$, $2$,
respectively. The Entangler is to implement the transformation,
\begin{equation}
\left\vert \psi^{^{\prime}}\right\rangle _{S}\left\vert +\right\rangle
_{a}\rightarrow\alpha^{^{\prime}}\left\vert HH\right\rangle _{0,a}%
+\beta^{^{\prime}}\left\vert VV\right\rangle _{1,a}+\gamma^{^{\prime}%
}\left\vert VV\right\rangle _{2,a},
\end{equation}
where the polarization of the ancilla single photon will be the same as that
of the single photon qutrit. In the Entangler, two qubus beams $\left\vert
\alpha\right\rangle \left\vert \alpha\right\rangle $ are introduced, and then
coupled to the corresponding photonic modes through the XPM processes. The XPM
pattern in Fig. 5 is that the first beam being coupled to $\left\vert
V\right\rangle $ modes on path 1, 2 and the $\left\vert H\right\rangle $ mode
of the ancilla photon, while the second beam to $\left\vert H\right\rangle $
mode on path 0 and the $\left\vert V\right\rangle $ mode of the ancilla
photon. Suppose the XPM phase shifts induced by the couplings are all $\theta
$. As the result, we will transform the total system to
\begin{align}
&  \frac{1}{\sqrt{2}}\left(  \alpha^{^{\prime}}\left\vert HH\right\rangle
_{0,a}+\beta^{^{\prime}}\left\vert VV\right\rangle _{1,a}+\gamma^{^{\prime}%
}\left\vert VV\right\rangle _{2,a}\right)  \left\vert \alpha e^{i\theta
}\right\rangle \left\vert \alpha e^{i\theta}\right\rangle \nonumber\\
&  +\frac{1}{\sqrt{2}}\alpha^{^{\prime}}\left\vert HV\right\rangle
_{0,a}\left\vert \alpha\right\rangle \left\vert \alpha e^{i2\theta
}\right\rangle \nonumber\\
&  +\frac{1}{\sqrt{2}}\left(  \beta^{^{\prime}}\left\vert VH\right\rangle
_{1,a}+\gamma^{^{\prime}}\left\vert VH\right\rangle _{2,a}\right)  \left\vert
\alpha e^{i2\theta}\right\rangle \left\vert \alpha\right\rangle .
\end{align}
After that, a phase shifter of $-\theta$ is respectively applied to two qubus
beams, and then one more 50:50 BS implements the transformation $\left\vert
\alpha_{1}\right\rangle \left\vert \alpha_{2}\right\rangle \rightarrow
\left\vert \frac{\alpha_{1}-\alpha_{2}}{\sqrt{2}}\right\rangle \left\vert
\frac{\alpha_{1}+\alpha_{2}}{\sqrt{2}}\right\rangle $ of the coherent-state
components. The state of the total system will be therefore transformed to
\begin{align}
&  \frac{1}{\sqrt{2}}\left(  \alpha^{^{\prime}}\left\vert HH\right\rangle
_{0,a}+\beta^{^{\prime}}\left\vert VV\right\rangle _{1,a}+\gamma^{^{\prime}%
}\left\vert VV\right\rangle _{2,a}\right)  \left\vert 0\right\rangle
\left\vert \sqrt{2}\alpha\right\rangle \nonumber\\
&  +\frac{1}{\sqrt{2}}\alpha^{^{\prime}}\left\vert HV\right\rangle
_{0,a}\left\vert -i\sqrt{2}\alpha\sin\theta\right\rangle \left\vert \sqrt
{2}\alpha\cos\theta\right\rangle \nonumber\\
&  +\frac{1}{\sqrt{2}}\left(  \beta^{^{\prime}}\left\vert VH\right\rangle
_{1,a}+\gamma^{^{\prime}}\left\vert VH\right\rangle _{2,a}\right)  \left\vert
i\sqrt{2}\alpha\sin\theta\right\rangle \left\vert \sqrt{2}\alpha\cos
\theta\right\rangle . \label{ki}%
\end{align}
The first coherent-state component in Eq. (\ref{ki}) is either vacuum or a cat
state (the superposition of $\left\vert \pm i\sqrt{2}\alpha\sin\theta
\right\rangle $ in the second piece). The target output could be therefore
obtained by the projection $\left\vert n\right\rangle \left\langle
n\right\vert $ on the first qubus beam. If $n=0$, we will obtain
\begin{equation}
\alpha^{^{\prime}}\left\vert HH\right\rangle _{0,a}+\beta^{^{\prime}%
}\left\vert VV\right\rangle _{1,a}+\gamma^{^{\prime}}\left\vert
VV\right\rangle _{2,a}, \label{io}%
\end{equation}
with the polarization of ancilla photon the same to the single photon qutrit.
If $n\neq0$, on the other hand, there will be the output
\begin{equation}
e^{-in\frac{\pi}{2}}\alpha^{^{\prime}}\left\vert HV\right\rangle
_{0,a}+e^{in\frac{\pi}{2}}\left(  \beta^{^{\prime}}\left\vert VV\right\rangle
_{1,a}+\gamma^{^{\prime}}\left\vert VV\right\rangle _{2,a}\right)  ,
\end{equation}
which can be transformed to the form in Eq. (\ref{io}) by a phase shift $\pi$
following the classically feed-forwarded measurement result $n$ and a
$\sigma_{x}$ operation on the ancilla photon.

Next, the second Entangler will be applied to the above output and another
ancilla single photon $\left\vert \pm\right\rangle _{b}$, after a $\sigma_{x}$
operation is performed on spatial mode $1$. In this Entangler, the transmitted
path for mode 1 is now active while the reflected port is only active in the
first Entangler. Similar to the first Entangler, the second implements the
transformation
\begin{align}
&  \left(  \alpha^{^{\prime}}\left\vert HH\right\rangle _{0,a}+\beta
^{^{\prime}}\left\vert HV\right\rangle _{1,a}+\gamma^{^{\prime}}\left\vert
VV\right\rangle _{2,a}\right)  \left\vert +\right\rangle _{b}\nonumber\\
&  \rightarrow\alpha^{^{\prime}}\left\vert HHH\right\rangle _{0,a,b}%
+\beta^{^{\prime}}\left\vert HVH\right\rangle _{1,a,b}+\gamma^{^{\prime}%
}\left\vert VVV\right\rangle _{2,a,b}.
\end{align}
Also we need to erase the path information of the first photon. We first
combine the modes 1 and 2 by a PBS, and then make them interfere with the mode
0 through a 50:50 BS to achieve the following state
\begin{align}
&  \frac{\alpha^{^{\prime}}}{\sqrt{2}}\left(  \left\vert H\right\rangle
_{3}+\left\vert H\right\rangle _{4}\right)  \left\vert HH\right\rangle
_{a,b}+\frac{\beta^{^{\prime}}}{\sqrt{2}}\left(  \left\vert H\right\rangle
_{3}-\left\vert H\right\rangle _{4}\right)  \left\vert VH\right\rangle
_{a,b}\nonumber\\
&  +\frac{\gamma^{^{\prime}}}{\sqrt{2}}\left(  \left\vert V\right\rangle
_{3}-\left\vert V\right\rangle _{4}\right)  \left\vert VV\right\rangle _{a,b}.
\end{align}
By two PBS$_{\pm}$, the state
\begin{align}
&  \frac{1}{2}\left(  \alpha^{^{\prime}}\left\vert HH\right\rangle
_{a,b}+\beta^{^{\prime}}\left\vert VH\right\rangle _{a,b}+\gamma^{^{\prime}%
}\left\vert VV\right\rangle _{a,b}\right)  \left\vert +\right\rangle
_{5}\nonumber\\
&  +\frac{1}{2}\left(  \alpha^{^{\prime}}\left\vert HH\right\rangle
_{a,b}+\beta^{^{\prime}}\left\vert VH\right\rangle _{a,b}-\gamma^{^{\prime}%
}\left\vert VV\right\rangle _{a,b}\right)  \left\vert -\right\rangle
_{6}\nonumber\\
&  +\frac{1}{2}\left(  \alpha^{^{\prime}}\left\vert HH\right\rangle
_{a,b}-\beta^{^{\prime}}\left\vert VH\right\rangle _{a,b}-\gamma^{^{\prime}%
}\left\vert VV\right\rangle _{a,b}\right)  \left\vert +\right\rangle
_{7}\nonumber\\
&  +\frac{1}{2}\left(  \alpha^{^{\prime}}\left\vert HH\right\rangle
_{a,b}-\beta^{^{\prime}}\left\vert VH\right\rangle _{a,b}+\gamma^{^{\prime}%
}\left\vert VV\right\rangle _{a,b}\right)  \left\vert -\right\rangle _{8}%
\end{align}
will be then obtained. With four detectors on path 5, 6, 7 and 8, as well as
the classical feedforward, the state
\begin{equation}
\alpha^{^{\prime}}\left\vert HH\right\rangle _{a,b}+\beta^{^{\prime}%
}\left\vert VH\right\rangle _{a,b}+\gamma^{^{\prime}}\left\vert
VV\right\rangle _{a,b},
\end{equation}
will be finally realized. The above processes could be deterministic.

The final step is to merge the two photons into the same spatial mode, which
could be simply realized by a BS and the following QND module. In this QND
module, a qubus beam $\left\vert \alpha\right\rangle $ will be coupled to one
of the output modes of BS$_{1}$. After that, the state in Eq. (35) plus the
qubus beam will evolve to%
\begin{align}
&  \frac{1}{\sqrt{2}}\left(  \alpha^{^{\prime}}\left\vert HH\right\rangle
_{1,1}+\frac{1}{\sqrt{2}}\beta^{^{\prime}}\left\vert VH\right\rangle
_{1,1}+\gamma^{^{\prime}}\left\vert VV\right\rangle _{1,1}\right)  \left\vert
\alpha e^{i2\theta}\right\rangle \nonumber\\
&  +\frac{1}{\sqrt{2}}\beta^{^{\prime}}\left(  \left\vert VH\right\rangle
_{1,2}+\left\vert VH\right\rangle _{2,1}\right)  \left\vert \alpha e^{i\theta
}\right\rangle \nonumber\\
&  +\frac{1}{\sqrt{2}}\left(  \alpha^{^{\prime}}\left\vert HH\right\rangle
_{2,2}+\frac{1}{\sqrt{2}}\beta^{^{\prime}}\left\vert VH\right\rangle
_{2,2}+\gamma^{^{\prime}}\left\vert VV\right\rangle _{2,2}\right)  \left\vert
\alpha\right\rangle .
\end{align}
Through the quadrature measurement $\left\vert X\right\rangle \left\langle
X\right\vert $, the following state
\begin{equation}
\alpha^{^{\prime}}\left\vert HH\right\rangle _{1,1}+\frac{1}{\sqrt{2}}%
\beta^{^{\prime}}\left\vert VH\right\rangle _{1,1}+\gamma^{^{\prime}%
}\left\vert VV\right\rangle _{1,1}\label{out-1}%
\end{equation}
or
\begin{equation}
\alpha^{^{\prime}}\left\vert HH\right\rangle _{2,2}+\frac{1}{\sqrt{2}}%
\beta^{^{\prime}}\left\vert VH\right\rangle _{2,2}+\gamma^{^{\prime}%
}\left\vert VV\right\rangle _{2,2}\label{out-3}%
\end{equation}
can be selected out, and the output with only one photon at each output port,
which picks up the phase shift $\theta$ in the XPM process, will be discarded.
The different coefficients between the mid term and the other two terms in
Eqs. (\ref{out-1}) and (\ref{out-3}) are caused by the Hong-Ou-Mandal (HOM)
interference effect on BS$_{1}$. In order to balance the coefficients, we
should use a 50:50 BS respectively on path 0 and 2 (see Fig. 5). After that we
could achieve the bi-photonic qutrit $\alpha^{^{\prime}}\left\vert
0\right\rangle _{3}+\beta^{^{\prime}}\left\vert 1\right\rangle _{3}%
+\gamma^{^{\prime}}\left\vert 2\right\rangle _{3}$ with the success
probability $\frac{1}{2}$, and then the total success probability for an
arbitrary unitary operation on biphoton qutrits will be $\frac{1}{6}%
\times\frac{1}{2}=\frac{1}{12}$.

\section{Discussion}

\label{sec4} We have presented two schemes for unitary operations on biphoton
qutrits, which are realized through bi-directional mapping between
polarization and spatially encoded photonic qutrits. Through the
bi-directional mapping any unitary operation $U(3)$ on bi-photonic qutrits can
be reduced to that on single photon qutrits. The linear optical scheme
succeeds with a small probability $1.17\times10^{-4}$, but it can be increased
to $1/12$ with weak cross-Kerr nonlinearity. The probabilistic nature of the
schemes is due to the two indistinguishable photons in the same
spatial-temporal modes. For example, at the last merging step in Fig. 5, the
probability to get the proper output state will be lowered by $1/2$ because of
the HOM interference.

Finally, we look at the feasibility of the schemes. The first scheme applies
common experimental tools such as linear optical circuits, coincidence
measurements, and detection with APDs. The difficulty in the implementation is
the accuracy for the numerous interferences between the photonic modes. The
additional requirement in the second scheme is the good performance of weak
cross-Kerr nonlinearity. The error in each XPM process can be effectively
eliminated under the condition $\alpha\theta\gg1$ \cite{Spiller}, which means
that the small XPM phase $\theta$ can be compensated by the large amplitude
$|\alpha|$ of the qubus or communication beams. The other advantage of the
scheme based on weak nonlinearity is the fewer ancilla photons---the ancilla
photons are only required in the inverse transformation. This could make the
experimental implementation more simplified.

\begin{acknowledgments}
Q. L. thanks Dr. Jian Li and Ru-Bing Yang for helpful discussions.
\end{acknowledgments}


\begin{thebibliography}{99}
\bibitem {note}If we adopt the notation $\left\vert 0\right\rangle
_{3^{\prime}}\equiv\left\vert HH\right\rangle $, $\left\vert 1\right\rangle
_{3^{\prime}} \equiv\frac{1}{\sqrt{2}}(\left\vert HV\right\rangle +\left\vert
VH\right\rangle )$, $\left\vert 2\right\rangle _{3^{\prime}} \equiv\left\vert
VV\right\rangle $ in \cite{Lanyon}, there will be no difference from the
results obtained in the present paper, since any input in this basis can be
represented as $|\psi\rangle_{in}+Perm|\psi\rangle_{in}$, where $Perm$ refers
to the permutation of two indistinguishable photons, and $|\psi\rangle
_{in}=\frac{\alpha}{2}|0\rangle_{3}+\frac{\beta}{\sqrt{2}}|1\rangle_{3}%
+\frac{\gamma}{2}|2\rangle_{3}$ is the qutrit defined in the logic basis of
Eq. (\ref{1}).
\end{thebibliography}
\end{document}